# Detecting Fake Escrow Websites using Rich Fraud Cues and Kernel Based Methods

Ahmed Abbasi and Hsinchun Chen
Artificial Intelligence Lab, Department of Management Information Systems
The University of Arizona, Tucson, Arizona 85721, USA
{aabbasi@email.arizona.edu, hchen@eller.arizona.edu}

**Abstract:** The ability to automatically detect fraudulent escrow websites is important in order to alleviate online auction fraud. Despite research on related topics, fake escrow website categorization has received little attention. In this study we evaluated the effectiveness of various features and techniques for detecting fake escrow websites. Our analysis included a rich set of features extracted from web page text, image, and link information. We also proposed a composite kernel tailored to represent the properties of fake websites, including content duplication and structural attributes. Experiments were conducted to assess the proposed features, techniques, and kernels on a test bed encompassing nearly 90,000 web pages derived from 410 legitimate and fake escrow sites. The combination of an extended feature set and the composite kernel attained over 98% accuracy when differentiating fake sites from real ones, using the support vector machines algorithm. The results suggest that automated web-based information systems for detecting fake escrow sites could be feasible and may be utilized as authentication mechanisms.

## 1. Introduction

Electronic markets have seen unprecedented growth in recent years. The lack of physical contact and prior interaction makes such places susceptible to opportunistic member behavior [12]. Consequently, fraud and deception are highly prevalent in electronic markets, particularly online auctions which account for 50% of internet fraud [4]. Although online escrow services (OES) are intended to serve as trusted third parties [6], their increased use has inevitably brought about the rise of escrow fraud. Escrow fraud is a variant of the popular "failure-to-ship" fraud; the seller creates a fake OES website and disappears after collecting the buyer's money [4]. Fraudulent OES sites often appear authentic and difficult to identify as fake by unsuspecting users [8]. While there has been recent work on spoof site detection, fake OES sites have received little attention despite their pervasiveness.

In this study we propose an approach designed to automatically identify fake escrow websites. The method incorporates a rich set of features and a classification technique based on a kernel specifically tailored towards representing the unique characteristics of these fake sites. The kernel enables effective categorization of fraudulent escrow web pages by considering content duplication patterns as well as website structural attributes. Results from this research serve two important purposes. Firstly, the study assesses the feasibility of automatic OES identification mechanisms that can help militate against the impact of Internet fraud stemming from fake escrow services. Secondly, evaluating different features and techniques for fake OES detection can provide insights into fraud patterns which may help educate Internet users that use trusted third party sites.

## 2. Related Work

The perceived effectiveness of OES plays a critical role in the level of online trust [12]. While OES are intended to offer security against the lack of identity trust, they themselves fall prey to similar concerns. It is often difficult for online traders to differentiate legitimate OES from fakes, making them susceptible to escrow fraud [4]. Automatic identification approaches could be utilized in a preemptive fashion, however they require considering the relevant features and techniques capable of fake OES detection.

While fake OES detection research has been scarce, there has been work on a related problem: web spam categorization. Web spam is the "injection of artificially created web pages in order to influence the results from search engines," [11, p. 83]. Analogous to fake OES sites, web spam typically uses automatic content generation techniques to mass produce fake web pages, resulting in many discernable content similarities. Prior web spam research has identified cues which may be applicable to fake OES sites [4]. These include feature categories pertaining to the following website segments: body text, HTML design, URL and anchor text, images, and site linkage and structure.

Lexical and syntactic features [2, 3] such as words per page, word n-gram frequencies, and misspellings are potential cues found in a web page's *body text* [11]. *HTML* source elements such as tag n-grams are helpful for identifying web page design similarities [14]. *URL and anchor text features* including URL length, number of dashes and digits, and certain URL suffixes provide useful fraud cues [4, 9]. Due to difficulties in indexing, many studies have ignored *image features* [9], however even simplistic features can facilitate the identification of duplicate images. *Linkage features*, including in-links and out-links have been used effectively for categorizing web pages based on similar topics [9].

One important difference between fake escrow sites and web spam is that web spam is intended to deceive search engines [11] while fake escrow sites are geared towards traders. In addition to text, fake OES must consider website design elements in order to aesthetically appeal to online buyers. Fake OES site detection likely involves the inclusion of multiple feature categories; however this introduces representational complexities for the classification techniques employed.

Detection of fake OES sites entails consideration of text, images, and structure. Two techniques used considerably in the past are support vector machine (SVM) and principal component analysis (PCA). SVM's effectiveness for categorization of style and images [10] makes it particularly suitable for fake OES detection. PCA has also been applied to text and image processing; using dimensionality reduction allows it to uncover important variation patterns [1]. A website contains many linked pages, with each comprised of body text, HTML, and images. Important fake OES detection characteristics include (1) the use of a rich feature set (2) stylistic patterns and duplication across sites (3) fraud cues inherent in website structure and linkage. A standard classifier can have difficulties handling such information. Two alternatives applicable with SVM and PCA are feature ensemble classifiers and kernel methods [10].

Feature ensemble classifiers are multiple classifiers built using different feature subsets. They have been effective for analysis of style and patterns [13]. While feature ensembles provide representational flexibility, feature set segmentation results in the loss of potentially informative feature interactions. Furthermore, a linear ensemble may still have problems considering structural information and accounting for content duplication tendencies. For complex structure information and problem-specific characteristics that cannot be described by standard feature vectors, kernel-based methods provide an effective alternative. Kernels can combine fraud cues inherent in text, linkage and image features along with website structural attributes [5]. Kolari et al. [7] noted that kernel functions could be useful for detecting fake websites, yet have not been explored. SVM kernels have been effective for classifying linked documents while Kernel PCA has been applied to image processing.

## 3. Research Questions

There is a need for fake OES detection methods given the difficulties people have in determining whether a particular site is legitimate [8]. However, it is unclear what features, techniques, and kernels will be effective. We are motivated to address the following questions:
- How will the use of all feature categories improve performance over individual categories?
- Which technique is better suited at differentiating fake escrow web sites from real ones?
- Can kernel methods outperform standard linear classifiers?

## 4. System Design

Our system is comprised of features, techniques, and kernels necessary for categorization of fake OES websites. The feature set utilized was comprised of features from the five aforementioned categories: body text, HTML, URL, image, and linkage (Table 1). Body text features include lexical measures and style markers described in prior research [1]. HTML tag n-grams were used for representing design style [14]. URL features included character and token level n-grams [11]. Image features consisted of pixel color frequencies. Link and structure features included page and site level in/out link attributes. Since n-gram features were selected using the information gain heuristic, their quantities were unknown apriori.

SVM and PCA were used since both have been applied to text and image categorization. In addition to the linear kernel, we propose a composite kernel (Figure 1) tailored to the properties of fake OES sites. The kernel compares each web page to every site (based on the 5 feature categories outlined above), and computes the average and maximum page-page similarity for each comparison in order to detect common patterns as well as content duplication, respectively. Additionally, the kernel function considers structural attributes of the pages being compared (e.g., the number of in/out links and page levels), allowing for a more accurate representation of website similarity.

**Table 1:** Fake OES Website Identification Feature Set

| Feature Group | Category | Quantity | Description/Examples |
|---|---|---|---|
| Body Text | Word/Char. Level Lexical | 10 | total words, total char., % char. per word |
| | Letter/Digit N-Grams | < 19,388 | count of letters and digits (e.g., a, at, ath, 12, 1) |
| | Word Length Dist. | 20 | frequency distribution of 1-20 letter words |
| | Vocabulary Richness | 8 | e.g., hapax legomena, Yule's K, Honore's H |
| | Punctuation/Special Char. | 29 | occurrence frequency of char. (e.g., !:;.?@#$%^) |
| | Function Words | 300 | frequency of function words (e.g., of, for, to) |
| | POS Tag N-Grams | varies | part-of-speech tags (e.g., NNP, NNP JJ) |
| | Document Structure | 64 | e.g., has greeting, has url, paragraphs, font colors |
| | Bag-of-word N-Grams | varies | e.g., "trusted", "third party", "trusted third" |
| | Misspelled Words | < 5,513 | e.g., "beleive", "thougth" |
| HTML | HTML tag N-Grams | varies | e.g., <HTML>, <HTML> <BODY> |
| URL | Character N-Grams | varies | e.g., a, at, ath, /, _, : |
| | Token N-Grams | varies | e.g. "spedition", "escrow", "trust", "online" |
| Image | Pixel Colors | 10,000 | frequency bins for pixel color ranges |
| | Image Structure | 40 | image extensions, heights, widths, file sizes |
| Link/Structure | Site and Page Linkage | 10 | site and page level relative/absolute in/out links |
| | Page Structure | 31 | page level, in/out link levels distribution |

Represent each page $a$ with the vector $x_a : \{Sim_{ave}(a, b_1), Sim_{max}(a, b_1), ..., Sim_{ave}(a, b_p), Sim_{max}(a, b_p)\}$

Where:

$$Sim_{ave}(a,b) = 1 - \left(\frac{1}{m}\sum_{k=1}^{m}\left(\left(\frac{|lv_a - lv_k|}{lv_a + lv_k}\right) * \left(\frac{|in_a - in_k|}{in_a + in_k}\right) * \left(\frac{|out_a - out_k|}{out_a + out_k}\right) * \left(\frac{1}{n}\sum_{i=1}^{n}\frac{|a_i - k_i|}{a_i + k_i}\right)\right)\right)$$

$$Sim_{max}(a,b) = \arg\max_{k \in \text{pages in site b}}\left(1 - \left(\left(\frac{|lv_a - lv_k|}{lv_a + lv_k}\right) * \left(\frac{|in_a - in_k|}{in_a + in_k}\right) * \left(\frac{|out_a - out_k|}{out_a + out_k}\right) * \left(\frac{1}{n}\sum_{i=1}^{n}\frac{|a_i - k_i|}{a_i + k_i}\right)\right)\right)$$

For:

$b \in p$ web sites in the training set; $k \in m$ pages in site b;

$lv_a, in_a,$ and $out_a$ are the page level and number of in/out links for page $a$;

$a_1, a_2, ... a_n$ and $k_1, k_2, ... k_n$ are page a and k's feature vectors;

The similarity between two pages is defined as the normalized inner product between their two vectors $x_1$ and $x_2$:

$$K(x_1, x_2) = \frac{\langle x_1, x_2 \rangle}{\sqrt{\langle x_1, x_1 \rangle \langle x_2, x_2 \rangle}}$$

**Figure 1:** Proposed Composite Kernel for Fake Escrow Websites

## 5. Hypotheses

We propose the following research hypotheses relating to the features, classification techniques, and kernels. Given the professional look of fake escrow websites [8], a rich set of features including text, image, and linkage attributes are likely necessary for effective detection. We suspect that the effectiveness of SVM on prior web spam research [5] will also translate into enhanced identification of fake escrow site. We also expect the composite kernel to better represent content similarities across fake escrow websites, resulting in improved performance over the linear kernel. The hypotheses regarding site level identification of OES are:

H1) The use of all features will outperform the use of any individual feature category.
H2) SVM will outperform PCA.
H3) The proposed composite kernel will outperform the linear kernel.

## 6. Evaluation

Experiments were conducted to evaluate the proposed feature set, classifiers, and kernels. This section encompasses a description of the experimental design, results, and hypotheses tests. We collected 350 fake OES and 60 real escrow sites over a three month period between 12/2006 and 2/2007. The fake OES URLs were taken from two online databases that post the addresses for verified fraudulent escrow sites: Escrow Fraud Prevention (http://escrow-fraud.com) and The Artists Against 4-1-9 (http://wiki.aa419.org). Since fake OES sites are often shut down or abandoned after they have been used, these sites typically have a short life span. In order to effectively collect these websites, we developed a web spidering program that monitored the online databases and collected newly posted URLs daily. Table 2 below shows the summary statistics for our test bed.

**Table 2:** Escrow Website Test Bed

| Category | # of Sites | # of Pages | # of Images | Pages Per Site | Images Per Site |
|---|---|---|---|---|---|
| Real OES Sites | 60 | 19,812 | 6,653 | 330.20 | 110.88 |
| Fake OES Sites | 350 | 69,684 | 29,764 | 199.10 | 85.04 |

The experimental design included 6 feature sets (5 individual categories and all), two techniques and two kernels, resulting in 24 experimental conditions. We ran 50 bootstrap instances for each condition, in which 50 real and 50 fake escrow sites were randomly selected each run. All web pages from the selected 100 sites were used as instances for that run and evaluated using 10-fold cross validation. PCA was run using the Kaiser-Guttman stopping rule which selects the $n$ eigenvectors with an eigenvalue greater than 1. Each test instance was assigned to the class with the lower average distance between its training points and the test point, across the $n$-dimensions [1]. The following evaluation metric was used:

$$\text{Accuracy} = \frac{\text{Number of Sites with Greater than 50\% Web Pages Correctly Classified}}{\text{Total Number of Websites}}$$

Table 3 and Figure 2 show the experimental results for the various conditions. Most features performed well, with accuracies over 95%. Using only image features attained up to 87% accuracy, suggesting that image duplication is pervasive in fake OES sites. The use of all features outperformed the individual feature categories with the linear ensemble and composite kernel, typically improving accuracy by 2%-10%. This supports the notion that OES fraud cues occur across feature categories. SVM outperformed PCA for most features except images. PCA's dimensionality reduction was useful for discerning image patterns [10]. With respect to kernel representations, kernel PCA outperformed linear PCA on all feature sets by a wide margin. For SVM, the linear kernel had enhanced performance over the composite kernel on the HTML attributes while the composite kernel outperformed linear SVM on all other feature sets.

**Table 3:** Average Classification Accuracy (%) across 50 Bootstrap Runs

| Technique and Kernel | Feature Set | | | | | |
|---|---|---|---|---|---|---|
| | Body Text | HTML | URL | Image | Link | All |
| Linear SVM | 97.68 | **97.80** | 94.36 | 76.04 | 93.92 | 97.96* |
| Kernel SVM | **97.86** | 97.74 | **97.72** | **83.14** | **95.69** | **98.44** |
| PCA | 71.50 | 72.34 | 70.66 | 83.81 | 70.90 | 81.86* |
| Kernel PCA | 82.15 | 80.14 | 82.48 | 87.65 | 76.33 | 86.93 |

*Linear Ensemble with 5 Classifiers

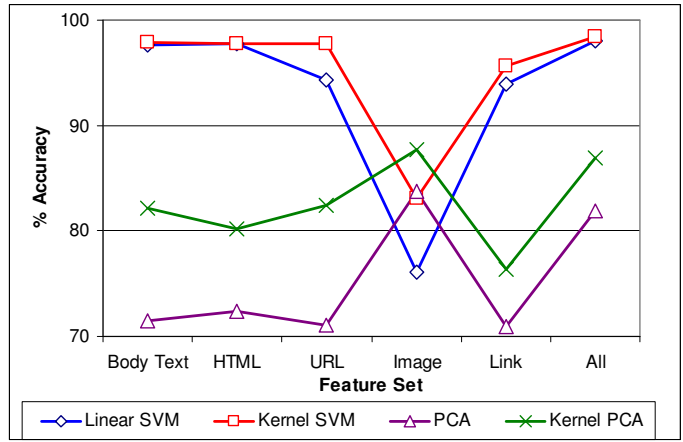

**Figure 2:** Performance for Various Combinations of Features, Techniques, and Kernels

We conducted pair wise t-tests on the 50 bootstrap runs. Given the large number of test conditions, a Bonferroni correction was performed in order to avoid spurious positive results. Only p-values less than 0.0004 were considered significant at alpha = 0.01. *H1. Features (All vs. Individual Categories):* The use of all features significantly outperformed individual feature sets using SVM and PCA with the linear and composite kernels. P-values on pair wise t-tests for all conditions were less than 0.0001 (n = 50). *H2. Techniques (SVM vs. PCA):* SVM significantly outperformed PCA for most feature sets (p-values < 0.0001), except image attributes, with the linear and composite kernel. *H3. Kernels (Linear vs. Composite Kernel):* The composite kernel significantly outperformed the linear approaches for all feature sets (11 out of 12 conditions) except for SVM when using HTML.

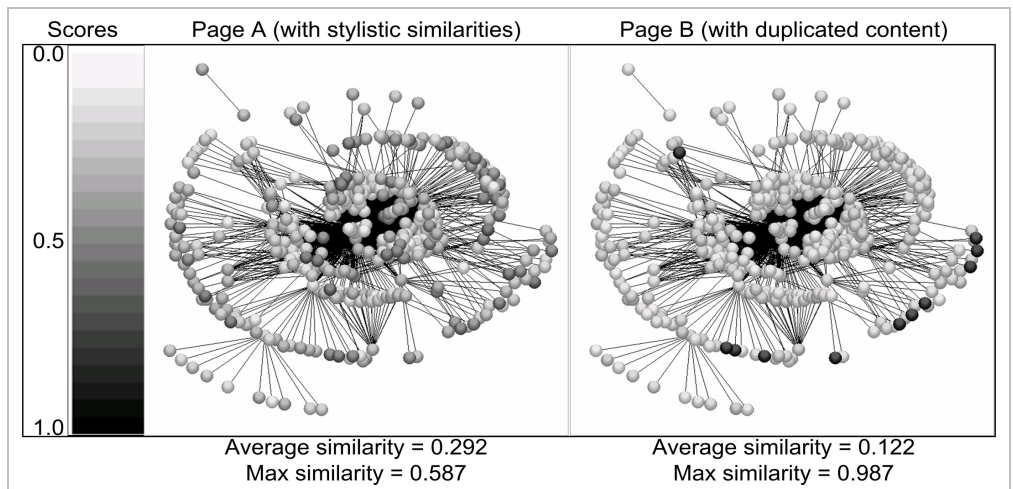

**Figure 3:** Similarities for Two Fake Pages Compared against Fraudulent OES Website www.bssew.com

The improved performance of the composite kernel was largely attributable to its use of both average and max similarity. Figure 3 depicts the similarity for two pages (A and B) taken from a fake OES site compared against all web pages from the fraudulent site www.bssew.com. The two graphs are the site maps for www.bssew.com, with each web page represented with a node, and lines between nodes indicating linkage. Each node's darkness indicates the level of similarity between that particular www.bssew.com page and pages A and B. The similarity scores are taken from the composite kernel, when using all features with SVM. Page A shares stylistic patterns with many pages in www.bssew.com, resulting in a high average similarity (evidenced by the predominance of gray page nodes). In contrast, Page B shares duplicated content with some pages in www.bssew.com, but little similarity with other pages (indicated by the presence of a few black page nodes and many lighter ones). The example illustrates how the composite kernel can effectively categorize unknown escrow website pages sharing stylistic tendencies or content duplication with other fake OES sites.

## 7. Conclusions

In this study we evaluated the effectiveness of automated approaches for fake OES website identification. The use of the proposed composite kernel coupled with a rich feature set and the SVM classifier was able to effectively identify fake OES sites with great accuracy. In addition to assessing the feasibility of automated detection, our analysis revealed several key findings. We observed that fake OES fraud cues are inherent in body text, HTML, URL, link, and image features and that the use of a rich attributes set is important for enhancing performance. We also found that fake OES sites share stylistic pattern and content duplication tendencies with other fake websites. These two properties can be successfully leveraged into an automated classifier by using a customized kernel function. The findings from this research have important implications for developers of web security information systems and Internet users engaging in online transactions. Knowledge of key fraud cues can be disseminated across the growing number of online resources and communities of practice that have emerged pertaining to Internet fraud. The knowledge gained in this study could also be applied towards the development of a browser plug-in to help protect against fake escrow websites.

## 8. References

[1] Abbasi, A. and Chen, H. "Writeprints: A Stylometric Approach to Identity-Level Identification and Similarity Detection in Cyberspace", ACM Transactions on Information Systems, 2008, forthcoming.
[2] Abbasi, A. and Chen, H. "Categorization and Analysis of Text in Computer Mediated Communication Archives using Visualization," In Proceedings of the ACM/IEEE Joint Conference on Digital Libraries, 2007, pp. 11-18.
[3] Abbasi, A. and Chen, H. "Applying Authorship Analysis to Arabic Web Content," In Proceedings of the IEEE International Conference on Intelligence and Security Informatics, 2005, pp. 183 – 197.
[4] Chua, C. E. H. and Wareham, J. "Fighting Internet Auction Fraud: An Assessment and Proposal," IEEE Computer, 2004, pp. 31–37.
[5] Drost, I. and Scheffer, T. "Thwarting the Nigritude Ultramarine: Learning to Identify Link Spam," In Proceedings of the European Conference on Machine Learning (ECML '05), 2005, pp. 96-107.
[6] Hu, X., Lin, Z., Whinston, A. B., and Zhang, H. "Hope or Hype: On the Viability of Escrow Services as Trusted Third Parties in Online Auction Environments," Information Systems Research, (15:3), 2004, pp. 236-249.
[7] Kolari, P., Finin, T., and Joshi, A. "SVMs for the Blogosphere: Blog Identification and Splog Detection," In AAAI Symposium on Computational Approaches to Analysing Weblogs, 2006.
[8] MacInnes, I., Damani, M. and Laska, J. "Electronic Commerce Fraud: Towards an Understanding of the Phenomenon," In Proceedings of the Hawaii International Conference on Systems Sciences (HICSS '05), 2005.
[9] Menczer, F., Pant, G., and Srinivasan, M. E. "Topical Web Crawlers: Evaluating Adaptive Algorithms," ACM Transactions on Internet Technology, (4:4), 2004, pp. 378-419.
[10] Muller, K., Mika, S., Ratsch, G., Tsuda, K., and Scholkopf, B. "An Introduction to Kernel-based Learning Algorithms," IEEE Transactions on Neural Networks, (12:2), 2001, pp. 181-201.
[11] Ntoulas, A., Najork, M., Manasse, M., and Fetterly, D. "Detecting Spam Web Pages through Content Analysis," In Proceedings of the International World Wide Web Conference (WWW '06), 2006, pp. 83-92.
[12] Pavlou, P. A. and Gefen, D. "Building Effective Online Marketplaces with Institution-Based Trust," Information Systems Research, (15:1), 2004, pp. 37-59.
[13] Stamatatos, E., and Widmer, G. "Music Performer Recognition Using an Ensemble of Simple Classifiers," In Proceedings of the 15th European Conference on Artificial Intelligence, 2002.
[14] Urvoy, T., Lavergne, T., and Filoche, P. "Tracking Web Spam with Hidden Style Similarity," In Proceedings of the 2nd International Workshop on Adversarial Information Retrieval on the Web (AIRWeb), 2006.